\documentclass[preprintnumbers,amsmath,amssymb]{revtex4} 

\usepackage{graphicx}
\usepackage{dcolumn}
\usepackage{bm}
\usepackage{fancyhdr, lastpage} 

\begin{document}

\title{Concentration profiles for fine and coarse sediments suspended by waves over ripples: An analytical study with the 1-DV gradient diffusion model}

\small To cite this article: Absi R. (2010) Concentration profiles for fine and coarse sediments suspended by waves over ripples: An analytical study with the 1-DV gradient diffusion model, \textit{Advances in Water Resources}, Elsevier, \textbf{33}(4), 411-418. 

\author{Rafik Absi}
  \email{rafik.absi@yahoo.fr}
\affiliation{EBI, Inst. Polytech. St-Louis, Cergy University, 32 Bd du Port, 95094 Cergy-Pontoise Cedex, France.}

\begin{abstract}
Field and laboratory measurements of suspended sediments over wave ripples show, for time-averaged concentration profiles in semi-log plots, a contrast between upward convex profiles for fine sand and upward concave profiles for coarse sand. 
Careful examination of experimental data for coarse sand shows a near-bed upward convex profile beneath the main upward concave profile. Available models fail to predict these two profiles for coarse sediments. 
The 1-DV gradient diffusion model predicts the main upward concave profile for coarse sediments thanks to a suitable $\beta(y)$-function (where $\beta$ is the inverse of the turbulent Schmidt number and $y$ is the distance from the bed). 
In order to predict the near-bed upward convex profile, an additional parameter $\alpha$ is needeed. 
This parameter could be related to settling velocity ($\alpha$ equal to inverse of dimensionless settling velocity) or to convective sediment entrainment process. 
The profiles are interpreted by a relation between second derivative of the logarithm of concentration and derivative of the product between sediment diffusivity and $\alpha$. 
\end{abstract}

\maketitle


\section{Introduction}


The description of sediment transport in the marine environment is very complex due to complex interactions between waves, currents, sediments and sea-bed. The understanding of the physical processes of sediment dynamics in this environment is of crucial importance for accurate predictions of near-shore sediment transport and sea bed changes. In pure wave or oscillatory flows, turbulence is restricted to a thin boundary layer. The presence of a current together with waves implies that the turbulence spreads to cover all the flow depth. 
Very close to the bottom, the turbulence is due to the oscillatory motion and predicted concentrations do not differ very much whether the turbulence due to the current is included or not \cite{Fredsoe}. 
Suspended sediment concentrations depend on the interaction between suspended sediments and the flow's turbulence.  
The feedback between sediment and turbulence has an important effect on sediment transport \cite{Conley}. 
When considering suspended sediment in oscillatory flows, it is necessary to distinguish between the sheet flow regime (Shields numbers $> 0.7$) and the ripple regime at lower shear stresses. 
In extreme wave conditions and/or shallow water the sea bottom is plane and sheet flow is the dominant transport mode \cite{Dohmen}. However, for moderate wave conditions and/or deep water, wave ripples can be formed on the sea bottom. 
If the ripples are relatively steep ($\eta_r / \lambda_r \geq 0.12$, where $\eta_r$ is the ripple height and $\lambda_r$ is the ripple wavelength), the mixing close to the bed is dominated by coherent, periodic vortex structures \cite{Nakato} \cite{Blondeaux} \cite{Nielsen} \cite{Thorne}. 
Above rippled beds, the mixing in the near-bed layer is dominated by the mechanism of vortex shedding which entrains sediments near flow reversal during each wave half cycle to several ripple heights \cite{vanderWerf}. 

Examination of concentration profiles for different sand sizes suspended in the same flow over wave ripples provides information about the involved mechanisms. 
Field and laboratory measurements of suspended sediments over wave ripples show a contrast between an upward convex concentration profiles, time-averaged in semi-log plots, for fine sand and an upward concave profiles for coarse sand \cite{McFetridge} \cite{Nielsen}. 
The classical gradient (Fickian) diffusion model allows a good description of fine sediment concentrations but fails to describe concentration profiles for coarse sand \cite{NielsenPF}. 
The finite-mixing-length model \cite{NielsenPF}, which is of similar generality to the Lagrangian Taylor's model, contains gradient (Fickian) diffusion model as a limiting case for $l_m /L \rightarrow 0$, where $l_m$ is the mixing length and $L$ the scale of the distribution under consideration. 
Even if the finite-mixing-length model improves the concentration profile $c(y)$ for coarse sand (where $y$ is the distance from the bed), a difference with experimental data remains. However, it is possible to improve this profile by a suitable turbulent mixing which is able to increase the upward concavity of $c(y)$ \cite{AbsiPF}. 
This is possible with a turbulent diffusivity $\varepsilon_s$ which takes into account the effect of grain size. 
The diffusivity of sediments $\varepsilon_s$ is related to the diffusivity of momentum, the eddy viscosity $\nu_t$, by a coefficient $\beta = \varepsilon_s / \nu_t$ (i.e., the inverse of the turbulent Schmidt number). 
In order to include the effect of sand grain size on $\varepsilon_s$, it is important to consider a realistic function for $\beta$ which is dependent on grain size and concentrations and therefore on the distance from the bed $y$. 

The value of $\beta$ has been the subject of much research. 
Researchers found that $\beta$ approaches unity for fine sediments and deviates for coarse ones \cite{Dohmen} \cite{Graf}. 
In suspension flows over movable beds, experiments show that depth-averaged $\beta$-values are smaller than unity ($\beta < 1$) for flows without bed forms while they are larger than unity ($\beta > 1$) for flows with bed forms \cite{Graf}. 
In oscillatory flows, it is known that cycle-mean sediment diffusivity above ripples is significantly greater than the cycle-mean eddy viscosity, i.e. $\beta > 1$ \cite{Nielsen} \cite{Thorne}. This difference has not yet been clearly explained, but seems rest on the spatial-temporal correlation between suspended concentrations and vertical velocities in the flow field \cite{Thorne}. 
For 1-DV models, the value of $\beta$ was suggested empirically a constant equal to about $4$ for rippled beds \cite{Nielsen} \cite{Thorne}. 
However, $\beta$ should depend on grain size and concentrations and should be therefore $y$-dependent.

Careful examination of measured concentrations for coarse sand shows a near-bed upward convex profile beneath the main upward concave profile. 
This near-bed profile was related to particles settling velocity which decreases near the bottom for high concentrations. Experiments have demonstrated that particle settling velocities are lower at higher concentrations. This behavior is described by the well known Richardson and Zaki \cite{Richardson} equation which depends on an empirically determined exponent. This equation was adapted for natural marine sands by Baldock \textit{et al.} \cite{Baldock}. 

The aim of this study is to: 1) improve the prediction of concentration profiles by the gradient (Fickian) diffusion model; 2) provide an analytical interpretation of the main upward concave concentration profile and the near-bed upward convex profile for coarse sand.  
This interpretation requires a relation between sediment diffusivity, particles settling velocity and sediment concentrations. 
The following Section presents gradient (Fickian) diffusion and finite-mixing-length modeling of suspended sediment concentrations. The effect of sand grain size on sediment diffusivity and concentration profiles will be considered in Section 3. Section 4 presents possible effects of settling velocity or convective sediment entrainment process on the near-bed upward convex profile for coarse sand.

\section{Gradient (Fickian) diffusion and finite-mixing-length modeling of suspended sediment concentrations}

Time-averaged concentrations of suspended sediment result from the balance between an upward mixing flux $\vec{q_m}$ and a downward settling flux $\vec{q_s} = c(y) \vec{\omega_s}$ as 

\begin{eqnarray}
q_m - c(y) \omega_s = 0 
\label{qmqs}
\end{eqnarray}

where $\omega_s$ is the particle settling velocity. 

\subsection{Classical gradient (Fickian) diffusion model} 

The gradient or Fickian diffusion model assumes that the mixing flux is proportional to
the concentration gradient 

\begin{eqnarray}
q_m = - \varepsilon_s \frac{d \: c} {d \: y}
\label{qmFick}
\end{eqnarray}

where $\varepsilon_s$ is the sediment diffusivity. 
Equations (\ref{qmqs}) and (\ref{qmFick}) allow to write the classical diffusion equation

\begin{eqnarray}
\frac{d \: c} {d \: y} = - \frac{\omega_s} {\varepsilon_s} \: c 
\label{cFick}
\end{eqnarray}

Equation (\ref{cFick}) needs the sediment diffusivity $\varepsilon_s$ which is related to the eddy viscosity $\nu_t$ by the parameter $\beta$ (i.e., the inverse of the turbulent Schmidt number) 
\begin{eqnarray} 
\varepsilon_s = \beta \: \nu_t 
\label{D_t}
\end{eqnarray} 
In this equation, $\beta$ describes the difference between diffusivity of momentum (diffusion of  a fluid ``particle") and diffusivity of sediment particles. 

\subsection{Finite-mixing-length model} 

In the finite-mixing-length model \cite{NielsenPF} the swapping of fluid parcels (including suspended sediment) between different levels generates a net vertical flux of momentum and suspended sediment. The sediment concentrations in a lower parcel and in an upper parcel are respectively $c(y-l_m/2)$ and $c(y+l_m/2)$. If the parcels travel vertically with equal and opposite velocities $\pm v_m$, the resulting flux is 

\begin{eqnarray}
q_m = v_m \: \left[ c\left( y - l_m/2 \right) - c\left( y + l_m/2 \right) \right]
\label{qmFML1}
\end{eqnarray}

where $v_m$ is the mixing velocity and $l_m$ the mixing length. The Taylor expansion of $c\left( y - l_m/2 \right)$ and  $c\left( y + l_m/2 \right)$ gives 

\begin{eqnarray}
q_m = - v_m l_m \left[ \frac{d \: c} {d \: y} + \frac{l_m^2 d^3 \: c} {24 d \: y^3} + ... \right]
\label{qmFML2}
\end{eqnarray}

Inserting Eq. (\ref{qmFML2}) into (\ref{qmqs}) 
and by including only the first two terms of the Taylor expansion in the brackets \cite{NielsenPF}, gives a third order ordinary differential equation (for more details see \cite{NielsenPF}) 

\begin{eqnarray}
\frac{l_m^2} {24} \: \frac{d^3 \: c} {d \: y^3} + \frac{d \: c} {d \: y} + \frac{\omega_s}{v_m l_m} c(y) = 0 
\label{qmFML}
\end{eqnarray}


Equations (\ref{qmFick}) and (\ref{qmFML2}) give 

\begin{eqnarray}
\varepsilon_s  = v_m l_m \left[ 1 + \frac{l_m^2} {24} \: \frac{\displaystyle \frac{d^3 \: c} {d \: y^3}} {\displaystyle \frac{d \: c} {d \: y}} + ... \right] 
\label{epsFick}
\end{eqnarray}

If the eddy viscosity $v_t$ is given by a similar expression to (\ref{epsFick}) with $c$ replaced by $u$, the parameter $\beta$ is determined by the ratio between the function in the brackets for each quantity 
or with $L_c^{-2} \approx (d^3 \: c / d \: y^3) / (d \: c / d \: y)$ and $L_u^{-2} \approx (d^3 \: u / d \: y^3) / (d \: u /  d \: y)$ as \cite{NielsenPF} 

\begin{eqnarray}
\beta  = \frac{\varepsilon_s}{\nu_t} 
= \frac{1 + \displaystyle \frac{l_m^2} {24} L_c^{-2} + ...}{1 + \displaystyle \frac{l_m^2} {24} L_u^{-2} + ...} 
= 1 + \frac{l_m^2} {24} \: \left( L_c^{-2} - L_u^{-2} \right) + ... 
\label{betaFML}
\end{eqnarray}

\subsection{Turbulent mixing} 

In this study we will use a simple algebraic eddy viscosity model given by 
\begin{eqnarray}
\nu_{t} = \alpha_1 \: \kappa \: u_* \: y \: e^{\displaystyle - C_1 \: \frac{y}{\delta}} 
\label{nut1P} 
\end{eqnarray}
where $\kappa$ is the karman constant ($=0.41$), $u_*$ the friction velocity, $\delta$ the boundary layer thickness and $\alpha_1$ and $C_1$ two parameters.

Eq. (\ref{nut1P}) was used in oscillatory boundary layers over rough flat beds by Hsu and Jan \cite{Hsu} and over wave ripples by Nielsen and Teakle \cite{NielsenPF}. 

Hsu and Jan \cite{Hsu} showed that the vertical profile of Eq. (\ref{nut1P}) is similar to the time-averaged $k$-$\varepsilon$ two-equation model. Predicted velocity profiles in oscillatory boundary layers showed very good agreement with experimental data \cite{Hsu}.

Nielsen and Teakle \cite{NielsenPF} used a similar eddy viscosity formulation given by 
\begin{eqnarray}
\nu_{t} = \lambda \: y \ v_{m0} \: e^{\left[\displaystyle - \frac{(y-y_0)}{L_v}\right]} 
\label{nut1}
\end{eqnarray}
where $v_{m0} = v_m(y=y_0)$ is the mixing velocity at level $y_0$; $\lambda$ and $L_v$ are two parameters. 
This model is based on a linearly increasing mixing length and an exponentially decreasing mixing velocity given respectively by 

\begin{eqnarray}
l_m = \lambda \: y \ \ \ \ \ \ and \ \ \ \ \ \ v_m = v_{m0} \: e^{\left[\displaystyle - \frac{(y-y_0)}{L_v}\right]}  
\label{lmvm}
\end{eqnarray}

The exponentially decreasing mixing velocity $v_m$ was chosen in agreement with turbulence measurements of du Toit and Sleath \cite{duToit} who observed that the turbulence intensity decayed exponentially upwards from a level $y_0$ very close to the ripple crest. Eq. (\ref{nut1}) reverts to Eq. (\ref{nut1P}) with $\alpha_1 \: \kappa \: u_* = \lambda \ v_{m0} \: e^{\displaystyle \left[\frac{y_0}{L_v}\right]}$ and $L_v = \displaystyle \frac{\delta}{C_1}$ \cite{AbsiPF}.

\subsection{Gradient (Fickian) diffusion versus finite-mixing-length model}

In order to analyze concentration profiles obtained by gradient (Fickian) diffusion and finite-mixing-length models, we use experimental data obtained by McFetridge and Nielsen \cite{McFetridge}. 

\subsubsection{Experiments of McFetridge and Nielsen} 

In these experiments \cite{McFetridge}, suspended sediments are due to non-breaking waves over rippled beds. 
In order to establish reliable time-averaged concentration profiles, an extensive series of trials under a single set of wave conditions was chosen rather than a limited number of trials under a variety of wave conditions. 
Natural beach sand was used, 
wave maker piston amplitude, period, and water depth were adjusted to maximize near bed velocities and therefore suspended sediment concentrations. The flow parameters are: wave period $T = 1.51 \: seconds$, mean wave height $H = 13 \: cm$, maximum near-bed flow velocity $U_{0m} = 27.8 \: cm/s$, near-bed flow semi-excursion $a = 6.68 \: cm$ and maximum return velocity at the bed $= 21.6 \: cm/s$, mean depth of flow $h = 30 \: cm$. 
The ripples which were produced were highly uniform and regular with a mean ripple height of $1.1 \: cm$, a mean ripple length of $7.8 \: cm$ and therefore a mean ripple steepness of $0.14$. 
Measured concentrations were obtained by sieving suction samples from different elevations above ripple crest \cite{McFetridge}. 

\subsubsection{Time-averaged concentration profiles} 

Fig.~(1) shows a comparison between time-averaged concentration profiles, $ln(c)$ versus linear $y$ plots, obtained with gradient (Fickian) diffusion and finite-mixing-length models and measurements \cite{McFetridge} for fine ($d = 0.06-0.11 \: mm$) and coarse sediments ($d = 0.42-0.60 \: mm$), where $d$ is grain diameter. Settling velocities are respectively $\omega_s=0.65 \: cm/s$ for fine and $\omega_s=6.1 \: cm/s$ for coarse sediments \cite{NielsenPF}. 

Predicted concentration profiles were obtained by solving differential equations (\ref{cFick}) and (\ref{qmFML}) with respectively (\ref{nut1}) and (\ref{lmvm}). We use the parameters proposed by Nielsen and Teakle \cite{NielsenPF}, namely $\lambda = 1$, $y_0 = 0.005 \: m$, $v_{m0} = 0.025 \: m/s$, $L_v = 0.022 \: m$. Concentration profiles (figure 1) are the same as profiles of figure (6 of \cite{NielsenPF}). Even if the upward convex profile for fine sand versus upward concave for coarse sand is reproduced with the finite-mixing-length model, it remains a difference with experimental data for coarse sand at the top of Fig.~(1) ($y$ between $0.04$ and $0.09 \: m$) which seems to become more important for $y > 0.09 \: m$. 
Since the aim of the study is to predict observed differences between concentration profiles for fine versus coarse sand suspended in the same flow, the turbulent flow's parameters were the same for fine and coarse sand (figure 1).

\begin{figure}
\includegraphics[width=10cm,height=10cm]{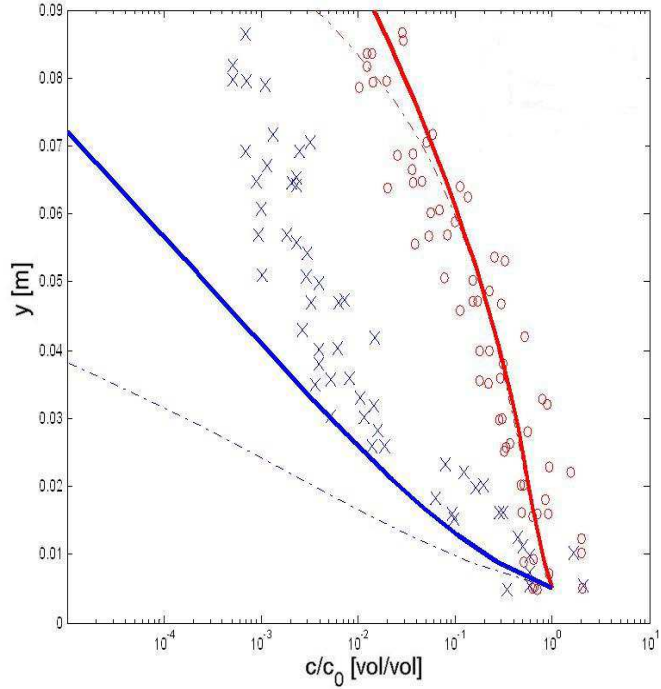}  
\caption{\label{fig:Figure1} Time-averaged concentration profiles over wave ripples: gradient (Fickian) diffusion model versus finite-mixing-length model. symbols, measurements \cite{McFetridge}, o, Fine sand ($\omega_s=0.65 \: cm/s$); $\times$, coarse sand ($\omega_s=6.1 \: cm/s$). Dash-dotted lines are solutions of gradient (Fickian) diffusion model (Eq. \ref{cFick}) while solid lines are solutions of finite-mixing-length model (Eq. \ref{qmFML}), $\lambda = 1$, $y_0 = 0.005 \: m$, $v_{m0} = 0.025 \: m/s$, $L_v = 0.022 \: m$. 
}
\end{figure}

\section{Effect of sediment diffusivity on concentration profiles}

In order to investigate the effect of sediment diffusivity on concentration profiles, we consider in this section 
a settling velocity constant over 
the boundary layer thickness, equal to $\omega_{st}$ the terminal settling velocity in an infinite fluid. 
In our study we will consider that $\beta$ approaches unity for fine sediments and deviates for coarse ones.

\subsection{Concentration profile for fine sediments} 

The assumption $\beta \approx 1$ for fine sediments allows us to write 
\begin{eqnarray}
\varepsilon_s \approx \nu_t = \alpha_1 \: \kappa \: u_* \: y \: e^{\displaystyle - C_1 \: \frac{y}{\delta}} 
\label{D_tF}
\end{eqnarray}

Careful examination of time-averaged concentration profile (Fig.~1) for fine sand ($\omega_s=0.65 \: cm/s$) shows 
a small near-bed upward concave profile followed by the main upward convex profile (Fig.~2.a). 
These profiles are associated to a sediment diffusivity $\varepsilon_s$ profile which increases then decreases (Fig. 2.b). 
The decreasing sediment diffusivity $\varepsilon_s$ allows an upward convex concentration $c(y)$ profile while an increasing sediment diffusivity $\varepsilon_s$ allows an upward concave $c(y)$ profile. 

Upward concavity/convexity of concentration profiles is, on $ln (c)$ versus linear $y$ plots, related to $\displaystyle \frac{d^2 \: ln \: c} {d \: y^2}$, while increasing/decreasing of $\varepsilon_s$ is related to $\displaystyle \frac{d \: \varepsilon_s} {d \: y}$. In order to understund this link, we need therefore a relation between $\displaystyle \frac{d^2 \: ln \: c} {d \: y^2}$ and $\displaystyle \frac{d \: \varepsilon_s} {d \: y}$. 

We are able to write 
\begin{eqnarray}
\frac{d \: ln \: c} {d \: y} =  \frac{1}{c}  \left( \frac{d \: c} {d \: y} \right) 
\label{dlnc}
\end{eqnarray}
and with Eq. (\ref{cFick}) 
\begin{eqnarray}
\frac{d \: ln \: c} {d \: y} =  - \frac{\omega_s} {\varepsilon_s}  
\label{dlnc2}
\end{eqnarray}

\begin{figure}
\includegraphics[width=6cm,height=6cm]{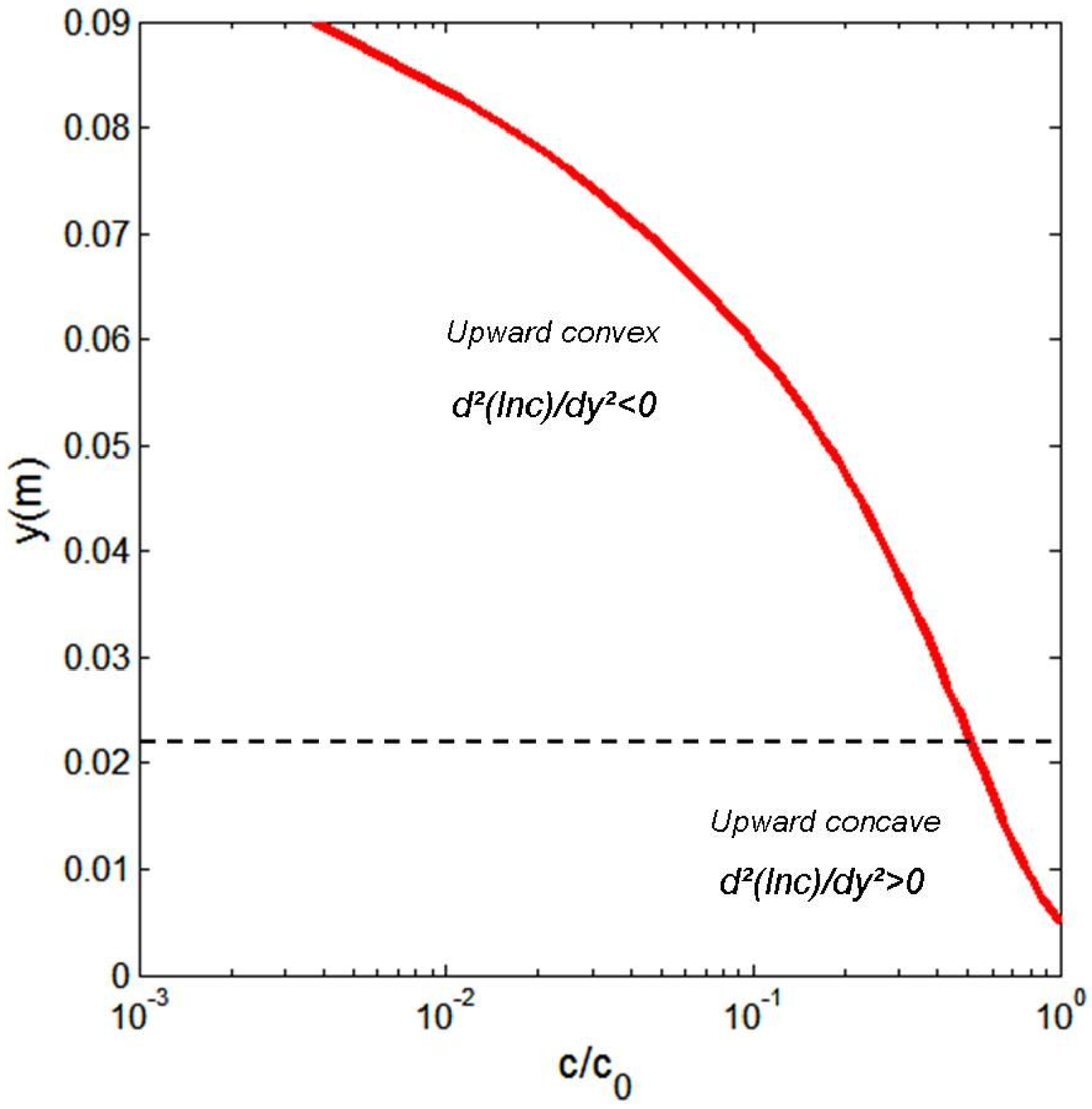} \\ 
\includegraphics[width=6cm,height=6cm]{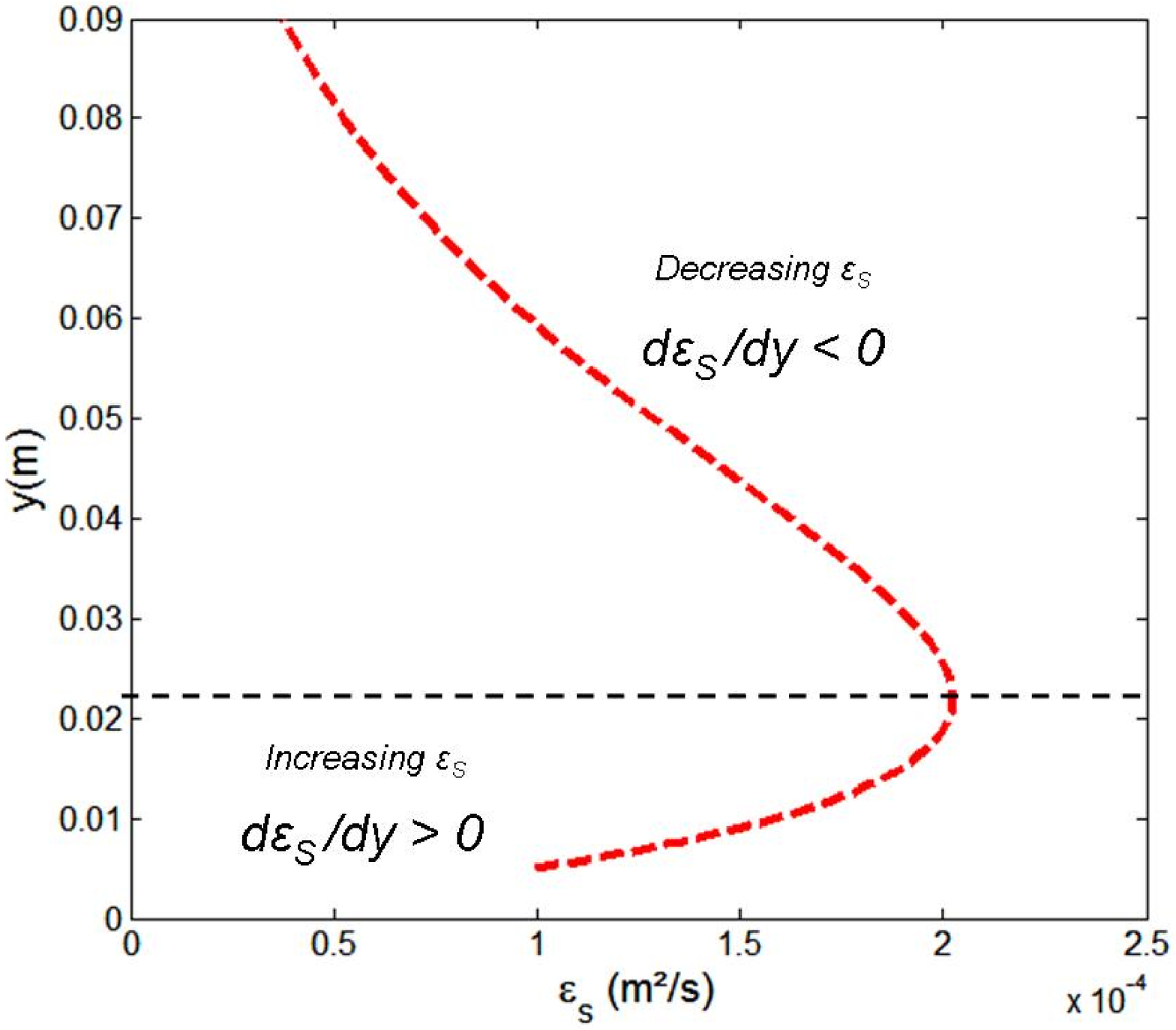}  
\caption{\label{fig:Figure2} Upward concavity/convexity of concetration profile for fine sand versus increasing/decreasing sediment difusivity. (a) Upward concavity/convexity of $c(y)$ (related to $\displaystyle \frac{d^2 \: ln \: c} {d \: y^2}$). (b)  increasing/decreasing $\varepsilon_s$ (related to $\displaystyle \frac{d \: \varepsilon_s} {d \: y}$). 
}
\end{figure}

The derivative of Eq. (\ref{cFick}) and Eq. (\ref{dlnc2}) are respectively  
\begin{eqnarray}
\frac{d^2 \: c} {d \: y^2} = \frac{\omega_s \: c} {\varepsilon_s^2} \: \left( \frac{d \: \varepsilon_s} {d \: y} + \omega_s \right) 
\label{d2cFick}
\end{eqnarray}

and

\begin{eqnarray}
\frac{d^2 \: ln \: c} {d \: y^2} =  \frac{1} {c} \left[ \frac{d^2 \: c} {d \: y^2} \right] - \frac{1} {c^2}  \left[ \frac{d \: c} {d \: y} \right]^2
\label{d2lnc}
\end{eqnarray}

Inserting Eq. (\ref{cFick}) and Eq. (\ref{d2cFick}) into Eq. (\ref{d2lnc}), we obtain 

\begin{eqnarray}
\frac{d^2 \: ln \: c} {d \: y^2} =  \frac{1} {c} \left[ \frac{\omega_s \: c} {\varepsilon_s^2} \: \left( \frac{d \: \varepsilon_s} {d \: y} + \omega_s \right) \right] - \frac{1} {c^2}  \left[ - \frac{\omega_s} {\varepsilon_s} \: c \right]^2
\label{d2lnc2}
\end{eqnarray}

After simplifying Eq. (\ref{d2lnc2}), we obtain 
\begin{eqnarray}
\frac{d^2 \: ln \: c} {d \: y^2} =  \frac{\omega_s} {\varepsilon_s^2} \: \frac{d \: \varepsilon_s} {d \: y} 
\label{d2lnc3}
\end{eqnarray}

Eq. (\ref{d2lnc3}) provides a link between $\displaystyle \frac{d^2 \: ln \: c} {d \: y^2}$ and $\displaystyle \frac{d \: \varepsilon_s} {d \: y}$, and therefore between upward concavity/convexity of concentration profiles and increasing/decreasing of sediment diffusivity profiles. 
Since $\displaystyle \frac{\omega_s} {\varepsilon_s^2}$ is always $>0$, $\displaystyle \frac{d^2 \: ln \: c} {d \: y^2}$ and $\displaystyle \frac{d \: \varepsilon_s} {d \: y}$ have the same sign and therefore increasing sediment diffusivity allows upward concave concentration profile, while decreasing sediment diffusivity allows an upward convex concentration profile (Fig.~2).

\subsection{Concentration profile for coarse sediments}

Due to grain size of coarse sediments, $\varepsilon_s$ becomes different from $\nu_t$. 
The difference in the shape of concentration profiles of fine and coarse sediments seems to be related to sand grain size which impacts sediment diffusivity through the parameter $\beta$. 

The analysis presented in the last section suggests that an increasing sediment diffusivity is needed to predict the upward concave concentration profile for coarse sand.

We write a $\beta(y)$-function as 
\begin{eqnarray}
\beta = \beta_b \: f_{\beta}(y) 
\label{beta1P}
\end{eqnarray}
where $\beta_b$ is the value of $\beta$ close to the bed. 

In order to obtain an increasing sediment diffusivity, we suggest 
\begin{eqnarray}
f_{\beta}(y) = e^{\displaystyle C_\beta \frac{y}{\delta}} 
\label{fbeta1P}
\end{eqnarray}
where $C_\beta$ is a coefficient which should depend on sand grain size.

No experimental data of $\beta(y)$ above wave orbital ripples are available to validate the proposed $\beta$-function. 
Available data from measurements, in steady and uniform open channel flows with Acoustic Particle Flux Profiler \cite{Graf}, show that $\beta$ increases with the distance from the bed, up to a given distance, indicating the higher the concentration, the lower are the $\beta$-values. The proposed $\beta$-function seems to be in agreement with these experimental data (fig. 3). 

It is possible to obtain a theroretical $\beta(y)$ profile, based on finite-mixing-length model, from Eq. (\ref{betaFML}) since $l_m$ is $y$-dependent. With a linear mixing length $l_m = \lambda \: y$, Eq. (\ref{betaFML}) gives 

\begin{eqnarray}
\beta = 1 + C_{\beta 2} \: y^2 
\label{betaFML2}
\end{eqnarray}

where $C_{\beta 2}$ is a coefficient ($C_{\beta 2}=(\lambda^2 / 24) \: \left( L_c^{-2} - L_u^{-2} \right)$). Fig. (3) shows that the profile of Eq. (\ref{betaFML2}) (dashed line) confirms the shape of Eq. (\ref{fbeta1P}) (solid line) and suggests that $\beta_b = 1$.

\begin{figure}
\includegraphics[width=6cm,height=6cm]{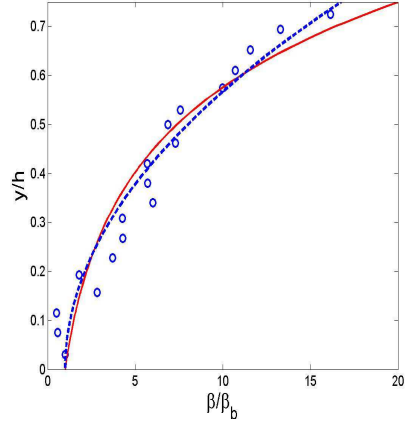}  
\caption{\label{fig:Figure3} Vertical distribution of dimensionless coefficient $\beta / \beta_b$. o,  experimental data \cite{Graf} (flow-sand condition Q$70S025$II). Curves, solid line: Eq. (\ref{fbeta1P}), dashed line: (Eq. \ref{betaFML2}) from finite-mixing-length model with $l_m = \lambda \: y$. 
}
\end{figure}

Using the $\beta(y)$-function (Eq. \ref{fbeta1P}) and eddy viscosity (Eq. \ref{nut1P}), the sediment diffusivity is therefore given by

\begin{eqnarray}
\varepsilon_s = (\beta_b \: \alpha_1 \: \kappa \: u_*) \: y \: e^{\displaystyle - (C_1 - C_\beta)  \frac{y}{\delta}} = \alpha_s \: y \: e^{\displaystyle -\frac{y}{\delta_s}}  
\label{Dt1P}
\end{eqnarray}
where $\alpha_s = \beta_b \: \alpha_1 \: \kappa \: u_*$ and $\displaystyle \delta_s = \frac{\delta}{(C_1 - C_\beta)}$ 
are the two parameters of sediment diffusivity.\\

\begin{figure}
\includegraphics[width=10cm,height=10cm]{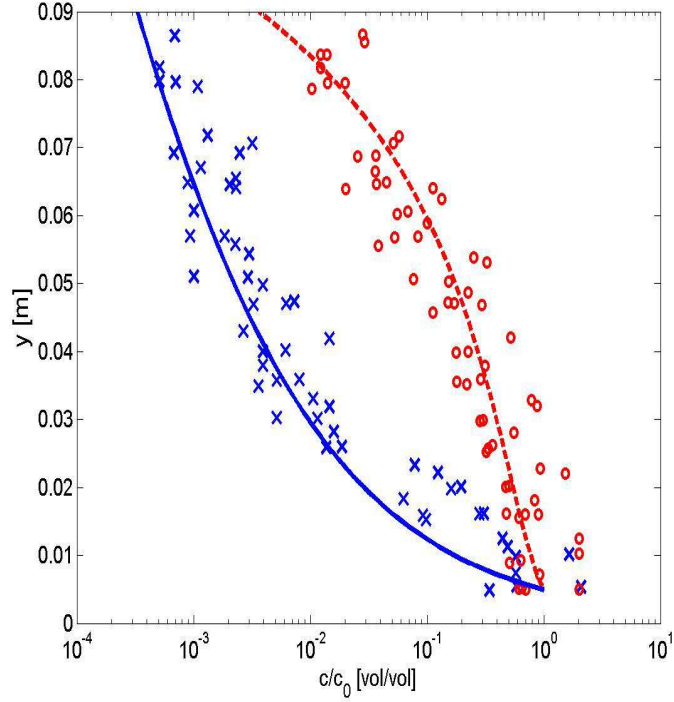}  
\caption{\label{fig:Figure4} Time-averaged concentration profiles over wave ripples: gradient (Fickian) diffusion model with proposed $\beta(y)$-function. symbols, measurements \cite{McFetridge}, o, Fine sand ($\omega_s = 0.65 \: cm/s$); $\times$, coarse sand ($\omega_s = 6.1 \: cm/s$). Curves, model with $\alpha_s = 0.025 \: m/s$; dashed line (fine sand) $\displaystyle \delta_s = L_v = \frac{\delta}{C_1} = 0.022 \: m$; solid line (coarse sand) $\displaystyle \delta_s = \frac{\delta}{(C_1 - C_\beta)} = 0.24 \: m$. 
}
\end{figure}

Results of time-averaged concentration profiles over wave ripples are presented in Fig.~(4) for fine ($\omega_s=0.65 \: cm/s$) and coarse ($\omega_s=6.1 \: cm/s$) sand. Comparisons with experimental data \cite{McFetridge} show good agreement. 
The upward convex profile for fine sand versus upward concave for the coarse sand is reproduced with the gradient (Fickian) diffusion model. 
The imperfection which was observed in figure (1) for coarse sand ($y$ between 0.04 and 0.09 m) is improved.

Our results show that the decreasing sediment diffusivity $\varepsilon_s$ for fine sand allows an upward convex concentration $c(y)$ profile while an increasing sediment diffusivity $\varepsilon_s$ for coarse sand allows an upward concave $c(y)$ profile (Fig.~4).\\

\section{The near-bed upward convex profile for coarse sand}

Careful examination of measured concentrations for coarse sand, in semi-log plots, shows a small near-bed upward convex concentration profile beneath the main upward concave profile (Fig.~5.a). 
Examination of other experimental data \cite{Dohmen} confirms the presence of near-bed upward convex profiles (Fig. 5.b).\\

\begin{figure}
\includegraphics[width=6cm,height=6cm]{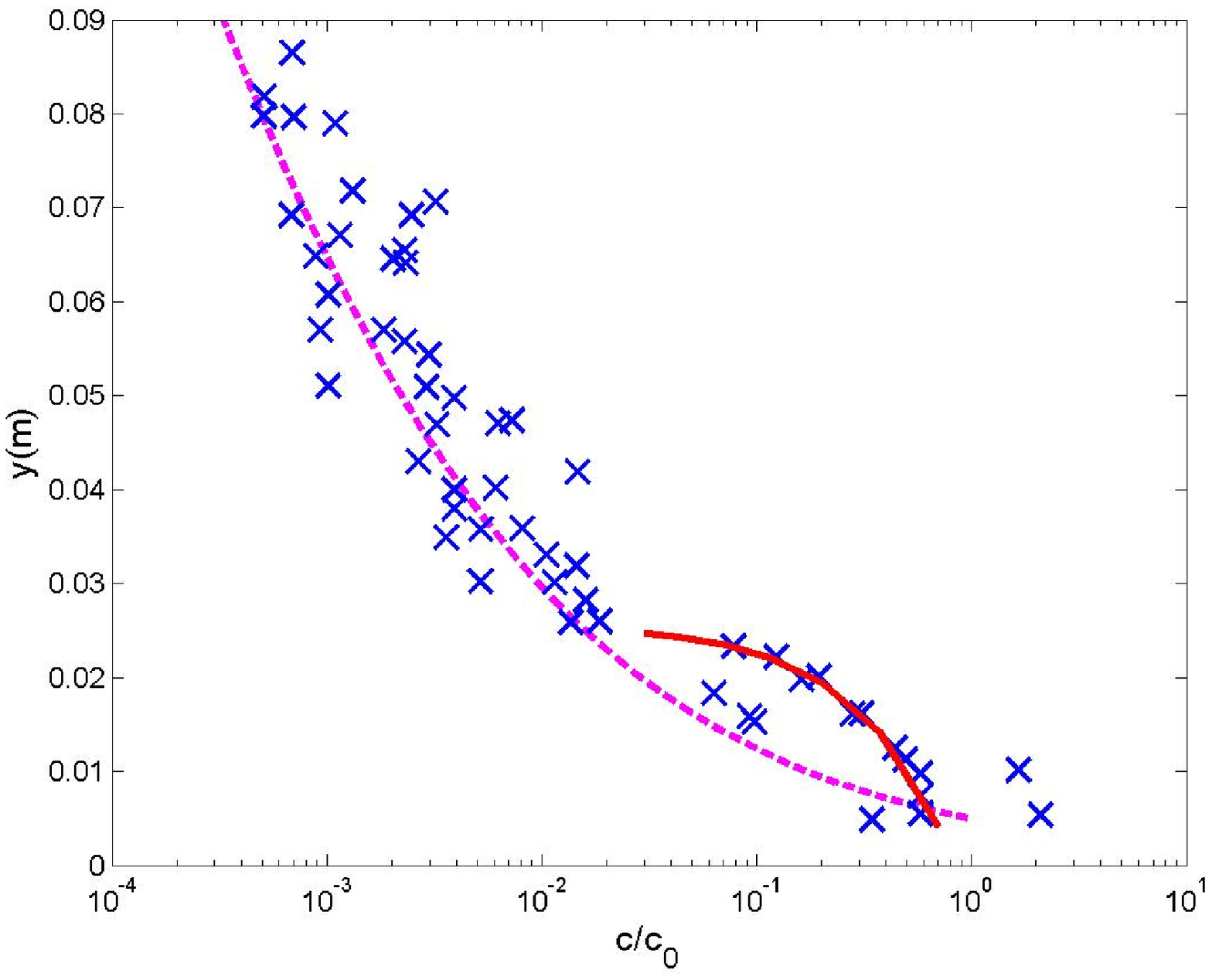} \\ 
\includegraphics[width=6cm,height=6cm]{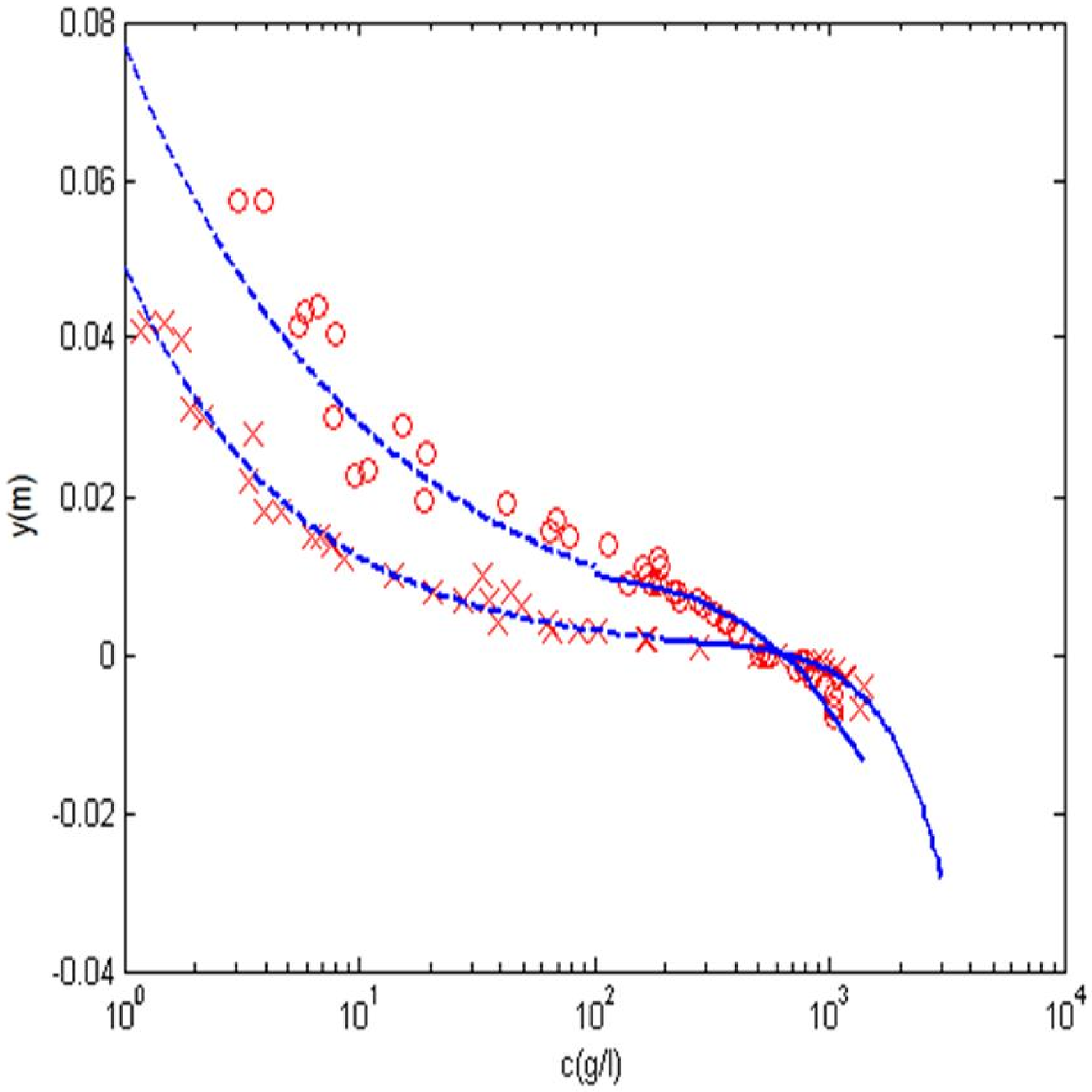}  
\caption{\label{fig:Figure5} Examination of experimental data; near-bed upward convex concentration profile. (a) Upward concave and near-bed upward convex concentration profiles (solid line) for coarse sand (data \cite{McFetridge}). 
(b) Time-averaged concentration profiles (data \cite{Dohmen}); o, Fine sand d=0,13 mm ;  $\times$, coarse sand d=0,32 mm; near-bed upward convex profiles (solid lines). 
}
\end{figure}

The 1-DV gradient diffusion model predicts the main upward concave profile for coarse sediments thanks to an adequate $\beta(y)$-function. The resulting sediment diffusivity $\varepsilon_s$ with a constant settling velocity ($\omega_s = \omega_{st}$) is unable to predict the near-bed upward convex profile. In order to predict this profile, we need an additional parameter. The sediment diffusivity should be modulated by an $y$-dependent parameter $\alpha$ which depends on sand grain size and decreases with $y$. If we consider a constant settling velocity ($\omega_s = \omega_{st}$), we write equation (\ref{cFick}) as 

\begin{eqnarray}
\frac{d \: c}{d \: y} = - \frac{\omega_s}{\alpha(y) \: \varepsilon_s(y)} \: c 
\label{cFick3} 
\end{eqnarray}

Eq. (\ref{d2lnc3}) becomes therefore 
\begin{eqnarray}
\frac{d^2 \: ln \: c} {d \: y^2} =  \frac{\omega_{s}} {\left(\alpha \: \varepsilon_s \right)^2} \: \frac{\displaystyle d \: \left(\alpha \: \varepsilon_s \right)} {d \: y} 
\label{d2lnc5}
\end{eqnarray}

As in the analysis presented in the last section, Eq. (\ref{d2lnc5}) provides a link between upward concavity/convexity of concentration profiles and increasing/decreasing of $\alpha \: \varepsilon_s$. 
Increasing $\alpha \: \varepsilon_s$ allows upward concave concentration profile, while decreasing $\alpha \: \varepsilon_s$ allows an upward convex concentration profile. 
Since $\varepsilon_s$ increases with $y$ for coarse sand, the near-bed upward convex concentration profile needs a decreasing $\alpha$.

\subsection{Possible effect of settling velocity} 

The near-bed upward convex profile could be related to settling velocity which decreases for high concentrations. 
Experiments have demonstrated that particle settling velocities are lower at higher concentrations. This behavior is given by the well known semi-empirical Richardson and Zaki \cite{Richardson} equation. We write Richardson and Zaki's equation as 
\begin{eqnarray}
\frac{\omega_s}{\omega_{st}} = \left(1 - \frac{c}{c_0}\right)^n 
\label{wsRZ}
\end{eqnarray}
where 
$c_0 = c(y=y_0)$ 
and $n$ is an empirically determined exponent dependent on the particle Reynolds number $R_t$ at $\omega_{st}$ and is constant for a particular particle. This exponent was determined experimentally as between $4.65$ and $2.4$ for increasing $R_t$. A review of empirical expressions for $n$ is given by Di Felice \cite{DiFelice}. These empirical values describing the variation of $n$ with $R_t$ were derived from fluidization and sedimentation experiments with spheres. Cheng \cite{Cheng} proposed expressions for $n$ which are functions of both Reynolds number and 
concentration. A method of determining $n$ for natural sands was proposed by Baldock \textit{et al.} \cite{Baldock} and was verified against laboratory fluidization data for marine sands. 
Settling velocity could be therefore not a constant on the vertical, and could depend on sediment grain size and concentrations and therefore on $y$. 
We write an $y$-dependent settling velocity, required by our analytical study, as 
\begin{eqnarray}
\omega_s(y) = \omega_{st} \ f_s(y) 
\label{ws1}
\end{eqnarray}
where $f_s(y)$ is a function which is equal to $1$ far from the bed where concentrations are small ($c\approx0$) and decreases near the bottom for high concentrations. 
In Section 3, we assumed that $\omega_s = \omega_{st} = constant$ and therefore $f_s = 1$. However, if this approximation is valid only far from the bed where concentrations are very small, the observed near-bed upward convex concentration profile for coarse sand could be related to $f_s(y)$. We need a suitable function for $f_s(y)$ in order to evaluate the effect on concentration profiles.

\subsubsection{$y$-dependent functions for settling velocity}

We propose two $y$-dependent functions for $\omega_s$. 
The first is based on a simple concentration profile (Fig.~6.a) given by 
\begin{eqnarray}
\frac{c(y)}{c_0} = exp\left(- \frac{y-y_0}{L_c}\right) 
\label{cexp} 
\end{eqnarray} 
Inserting equation (\ref{cexp}) into equation (\ref{wsRZ}), we obtain an $y$-dependent function for settling velocity 
\begin{eqnarray}
f_s(y) = \frac{\omega_s(y)}{\omega_{st}} = \left(1 - exp\left(- \frac{y-y_0}{L_c}\right)\right)^n 
\label{wsA1}
\end{eqnarray}

The second is an empirical function given by 
\begin{eqnarray}
f_s(y) = \frac{\omega_s(y)}{\omega_{st}} = \frac{1}{\displaystyle 1 + exp(- \frac{y-y_s}{h_s})} 
\label{wsE2}
\end{eqnarray}
where $y_s$ and $h_s$ are two parameters which should depend on concentrations and particles size. 
In order to validate these functions, data of settling velocity $w_s$ were derived from measured concentrations for coarse sand and Richardson and Zaki's equation. Data of $f_s(y)$ are therefore obtained from equation (\ref{wsRZ}) and measured $c/c_0$ values, as 
\begin{eqnarray}
f_s(y)_{exp} = \left(\frac{\omega_s(y)}{\omega_{st}}\right)_{exp} = \left(1 - \left(\frac{c(y)}{c_0}\right)_{exp} \right)^n 
\label{wsexp}
\end{eqnarray}

Experimental data obtained from equation (\ref{wsexp}) depend on the value of the exponent $n$. 
Fig.~(6.b) shows that the second function seems to be more accurate. 



\begin{figure}
\includegraphics[width=6cm,height=6cm]{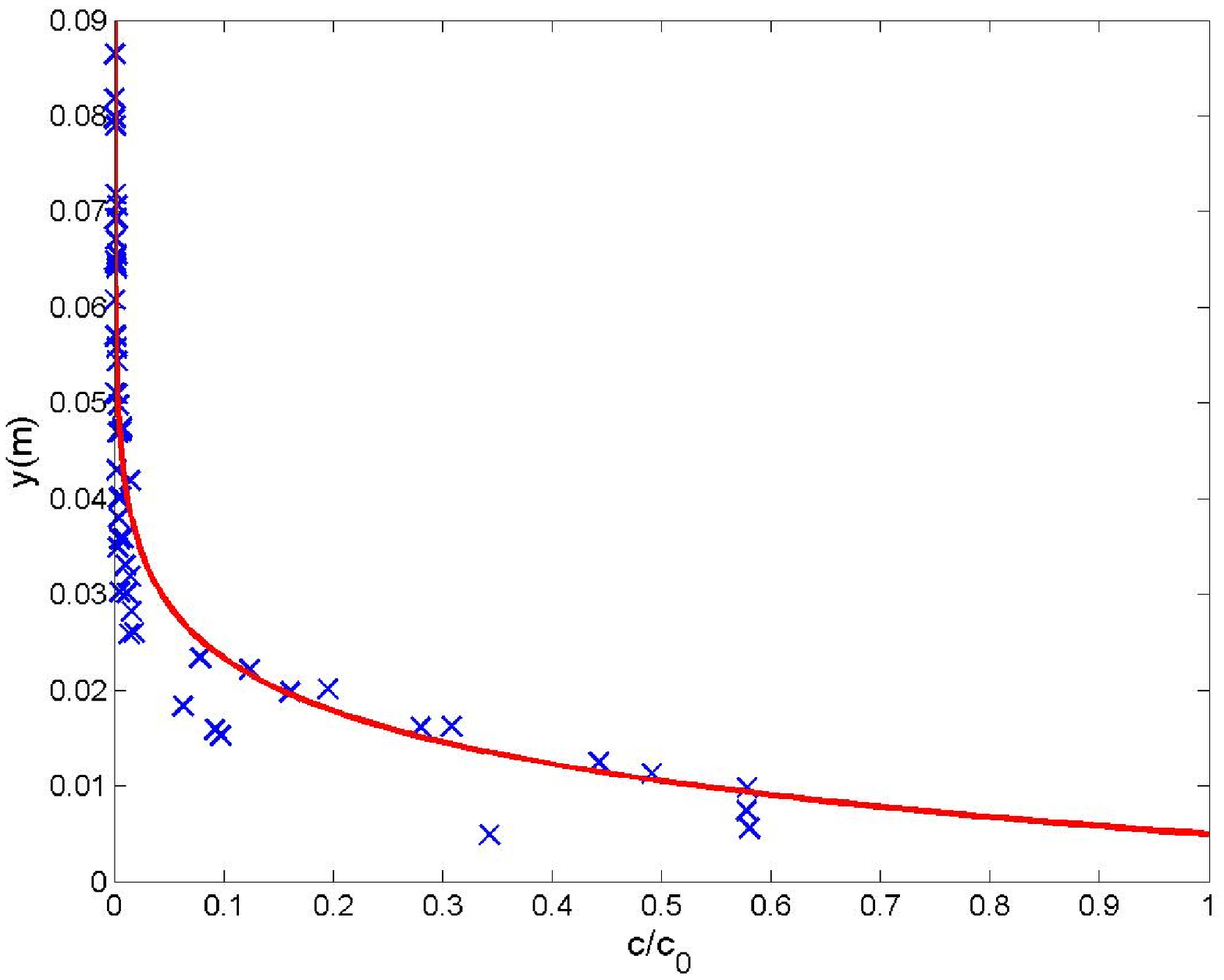} \\ 
\includegraphics[width=6cm,height=6cm]{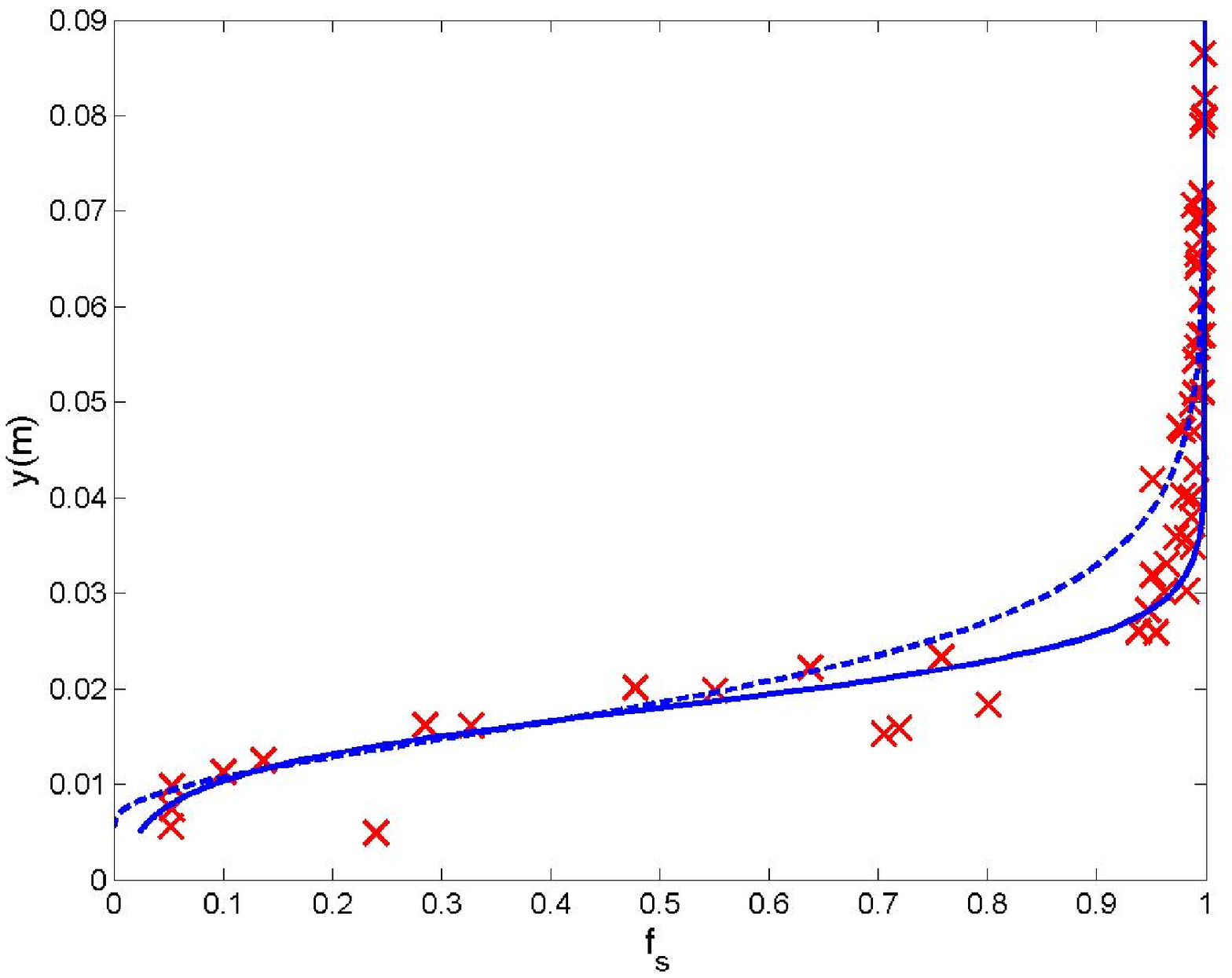}  
\caption{\label{fig:Figure6} (a) Concentration profile for coarse sand given by Eq. (\ref{cexp}), ($L_c \approx 0.008 \: m$); $\times$: measurements. (b) Settling velocity functions; $\times$: measurements from Eq. (\ref{wsexp}) with $n = 3.4$; dashed line: first settling velocity function (Eq. \ref{wsA1}, $L_c = 0.008 \: m$, $y_0 = 0.005 \: m$ and $n = 3.4$); solid line: second settling velocity function (Eq. \ref{wsE2}, $y_s = 0.018 \: m$ and $h_s = 0.0035 \: m$). 
}
\end{figure}

\subsubsection{Effect on concentration profiles}

Fig.~(7) allows to understand the possible effect of $f_s(y)$ on concentration profiles. 
In Section 3, $f_s$ was taken to be equal to $1$ and concentration profiles depend only on $\varepsilon_s$. 
Here, $f_s$ is equal to $1$ only far from the bottom where the concentration is small (Fig.~6.c). Fig.~(7.a) shows $1/f_s$ and Fig.~(7.b) $\varepsilon_s/f_s$. The shape of $\varepsilon_s/f_s$ allows to predict the near-bed upward convex concentration profile for coarse sand (Fig.~8).

\begin{figure}
\includegraphics[width=6cm,height=6cm]{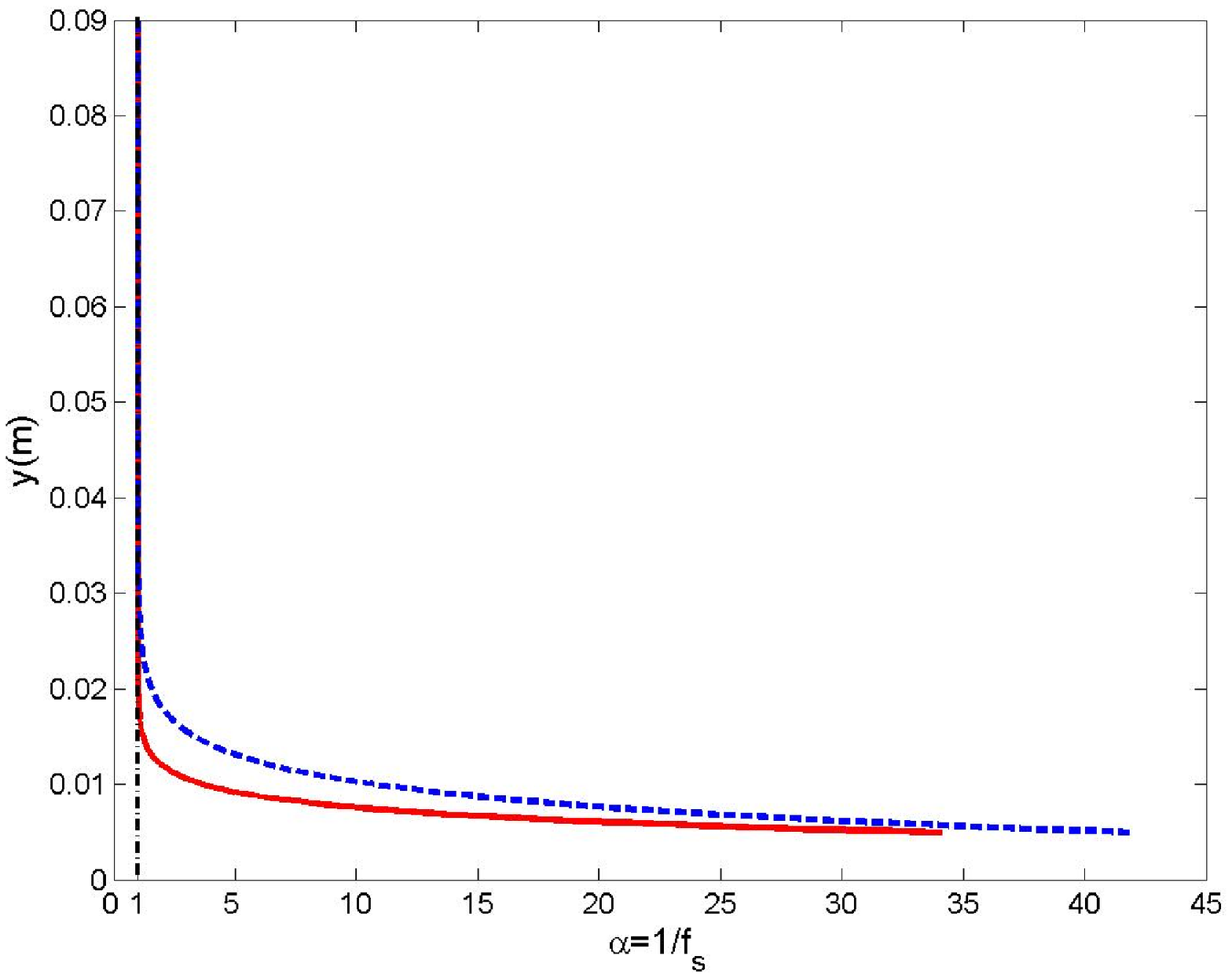} \\ 
\includegraphics[width=6cm,height=6cm]{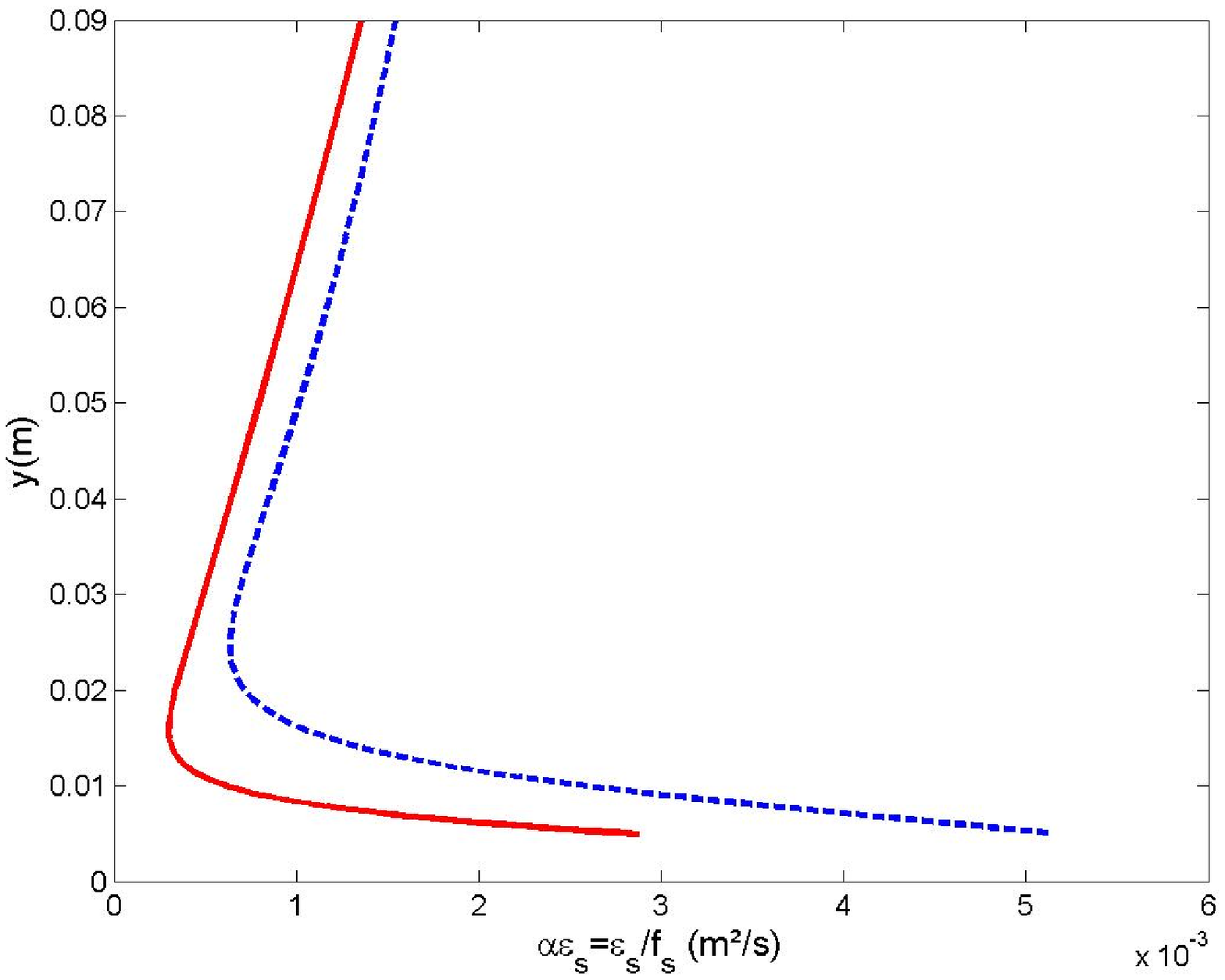}  
\caption{\label{fig:Figure7} Effect of dimensionless settling velocity profile $f_s(y)$ on predicted concentrations. (a) $1/f_s$. (b) $\varepsilon_s/f_s$. 
}
\end{figure}

Taking into account (Eq. \ref{ws1}), equation (\ref{cFick}) becomes 
\begin{eqnarray}
\frac{d \: c}{d \: y} = - \frac{\omega_s(y)}{\varepsilon_s(y)} \: c = - \frac{\omega_{st} \ f_s(y)}{\varepsilon_s(y)} \: c  = - \frac{\omega_{st}}{\displaystyle \frac{\varepsilon_s(y)}{f_s(y)}} \: c 
\label{cFick2}
\end{eqnarray} 

In Eq. (\ref{cFick2}), $\alpha=1/f_s$. 
Concentration profiles depend therefore on the ratio between sediment diffusivity and dimensionless settling velocity $\displaystyle \frac{\varepsilon_s(y)}{f_s(y)}$ and not on the sediment diffusivity $\varepsilon_s(y)$ alone.

Eq. (\ref{d2lnc3}) becomes therefore 
\begin{eqnarray}
\frac{d^2 \: ln \: c} {d \: y^2} =  \frac{\omega_{st}} {\displaystyle  \left( \frac{\varepsilon_s}{f_s} \right)^2} \: \frac{\displaystyle d \: \left( \frac{\varepsilon_s}{f_s} \right)} {d \: y} 
\label{d2lnc4}
\end{eqnarray}

We are able to generalize the result found in Section 3 by: upward concavity/convexity of concentration profiles is, on $ln(c)$ versus linear $y$ plots, related to increasing/decreasing of $\displaystyle \frac{\varepsilon_s(y)}{f_s(y)}$. 
Eq. (\ref{d2lnc4}) provides a link between $\displaystyle \frac{d^2 \: ln \: c} {d \: y^2}$ and $\displaystyle \frac{d \: \displaystyle \frac{\varepsilon_s(y)}{f_s(y)}} {d \: y}$, and therefore between upward concavity/convexity of concentration profiles and increasing/decreasing of $\displaystyle \frac{\varepsilon_s(y)}{f_s(y)}$. 
Since $\displaystyle \frac{\omega_{st}} {\displaystyle  \left( \frac{\varepsilon_s}{f_s} \right)^2}$ is always $>0$, $\displaystyle \frac{d^2 \: ln \: c} {d \: y^2}$ and $\displaystyle \frac{d \: \displaystyle \frac{\varepsilon_s(y)}{f_s(y)}} {d \: y}$ have the same sign and therefore increasing $\displaystyle \frac{\varepsilon_s(y)}{f_s(y)}$ allows upward concave concentration profile, while decreasing $\displaystyle \frac{\varepsilon_s(y)}{f_s(y)}$ allows an upward convex concentration profile. 
Decreasing profile of $\displaystyle \frac{\varepsilon_s(y)}{f_s(y)}$ in figure (7.b) between $0.005 m$ and $0.02 m$ allows to predict the near-bed upward convex $c(y)$ profile in figure (8).

Data of settling velocity $w_s$ were derived from measured concentrations for coarse sand and Richardson and Zaki's equation. 
The empirical function $f_s$ obtained from these data allows predicting the near-bed upward convex profile. 
However, the dimensionless settling velocity function $f_s$ decreases from $y=4 \: cm$ and at $y=2 \: cm$ the decreasing in settling velocity is of $50 \%$. This seems to be quite larger than would be expected and therefore seems to be outside the range of observed hindered settling.

\begin{figure}
\includegraphics[width=10cm,height=10cm]{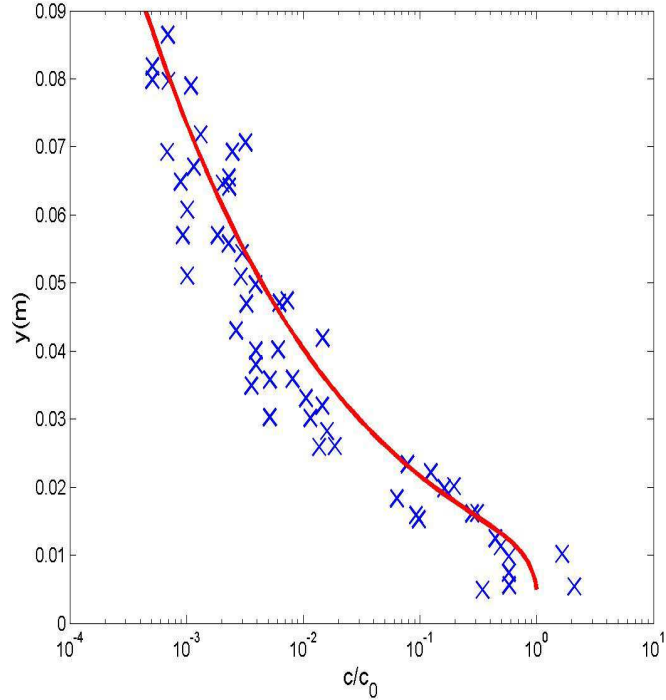}  
\caption{\label{fig:Figure8} Predicted time-averaged concentration profile for coarse sand ($\omega_s = 6.1 \: cm/s$). Curve, gradient (Fickian) diffusion Model with proposed $\beta(y)$ and $\alpha(y)$ ($=1/f_s(y)$) functions. 
$\times$, measurements \cite{McFetridge}. 
}
\end{figure}

\subsection{Possible effect of convective sediment entrainment process}


The process of vortex formation and shedding at flow reversal above ripples is a relatively coherent phenomenon. The associated convective sediment entrainment process may also be characterized as coherent, instead of a pure 
disorganized ``diffusive" process represented in the classical gradient diffusion model \cite{Thorne}. 
Above ripples, in a ripple-averaged sense, the convective term can 
dominate the upward sediment flux in the bottom part of the wave boundary layer \cite{Thorne2}. 
Nielsen \cite{Nielsen} indicated that both convective and diffusive mechanisms are involved in the entrainment processes. 
In the combined convection-diffusion formulation of Nielsen \cite{Nielsen}, the steady state advection-diffusion equation is given by 

\begin{eqnarray}
\omega_s \: c + \varepsilon_s(y) \: \frac{d \: c}{d \: y} - \omega_s \: c_0 \: F(y) = 0  
\label{cFick4} 
\end{eqnarray}

where $F(y)$ is a function describing the probability of a particle reaching height $y$ above the bed. The respective terms in Eq. (\ref{cFick4}) represent downward settling, upward diffusion, and upward convection 
\cite{Nielsen} \cite{LeeHanes} \cite{Thorne}. 
Eq. (\ref{cFick4}) reverts to Eq. (\ref{cFick3}) with 

\begin{eqnarray}
\alpha(y) = \frac{1}{1 - \left( c_0 / c \right) \: F(y)} 
\label{alpha2} 
\end{eqnarray}

In this equation, $\alpha$ represents the effect of upward convection. 
The near-bed upward convex profile of figure (8) could be predicted by using a parameter $\alpha(y) = 1 / f_s(y)$ based on  (Eq. \ref{wsE2}), $\alpha$ is given therefore by 

\begin{eqnarray}
\alpha(y) = \displaystyle 1 + exp(- \frac{y-y_s}{h_s}) 
\label{alpha3}
\end{eqnarray}



\section{Conclusion}

Field and laboratory measurements of suspended sediments over wave ripples show a contrast between an upward convex concentration profiles, time-averaged in semi-log plots, for fine sand and an upward concave profiles for coarse sand. 
Careful examination of experimental data for coarse sand shows a near-bed upward convex profile beneath the main upward concave profile. 

The 1-DV gradient diffusion model with a constant settling velocity ($w_s=cste$) and $\beta=1$, predicts concentration profile for fine sediments but fails for coarse sand. 

The 1-DV gradient diffusion model predicts the main upward concave profile for coarse sediments thanks to a suitable $\beta(y)$-function. The shape of this function was validated by the finite-mixing-length model. 

The model with the resulting sediment diffusivity $\varepsilon_s$ and a constant settling velocity is unable to predict the near-bed upward convex profile. In order to predict this profile, an additional parameter $\alpha$ is needed. The sediment diffusivity should be modulated by $\alpha$ which decreases with $y$. 

Data of settling velocity $w_s$ were derived from measured concentrations of coarse sand with Richardson and Zaki's equation. 
An empirical function for dimensionless settling velocity $f_s$ was obtained from these data. 
The 1-DV gradient diffusion model with the proposed function for $f_s$ predicts the near-bed upward convex profile. 
However, the dimensionless settling velocity $f_s$ decreases from $y=4 \: cm$ and at $y=2 \: cm$ the decreasing in settling velocity is of $50 \%$. This seems to be quite larger than would be expected and therefore seems to be outside the range of observed hindered settling. 

The used parameter $\alpha$ ($=1/f_s$) for the near-bed upward convex profile could be related to convective sediment entrainment process. 
Since above ripples the process of vortex formation and shedding at flow reversal is a relatively coherent phenomenon, both convective and diffusive mechanisms are involved in the entrainment processes. 
This seems to be in agreement with the combined convection-diffusion formulation of Nielsen \cite{Nielsen}. 


This study was based on experimental data of McFetridge and Nielsen and needs to be tested against other datasets in order to allow calibration for predictive purpose.




\begin{thebibliography}{breitestes Label}


\section*{References} 


\bibitem{AbsiPF} Absi R. Comment on 'Turbulent diffusion of momentum and suspended particles: A finite-mixing-length-theory'.  Phys Fluids 2005;17(7):079101. 
\bibitem{Baldock} Baldock TE, Tomkins MR, Nielsen P, Hughes MG. Settling velocity of sediments at high concentrations. Coast Eng 2004;51:91-100. 
\bibitem{Blondeaux} Blondeaux P, Vittori G. Vorticity dynamics in an oscillatory flow over a rippled bed, J Fluid Mech 1991;266:257-289. 
\bibitem{Cheng} Cheng NS. Effect of concentration on settling velocity of sediment particles. J Hydraul Eng, ASCE 1997;123(8): 728-731. 
\bibitem{Conley} Conley DC, Falchetti S, Lohmann IP, Brocchini M. The effects of flow stratification by non-cohesive sediment on transport in high-energy wave-driven flows. J Fluid Mech 2008;610:43-67. 
\bibitem{DiFelice} Di Felice R. Hydrodynamics of liquid fluidisation. Chem Eng Sci 2008;50:1213-1245. 
\bibitem{duToit} du Toit CG, Sleath JFA. Velocity measurements close to rippled beds in oscillatory flow. J Fluid Mech 1981;112:71-96. 
\bibitem{Dohmen} Dohmen-Janssen CM, Hassan WN, Ribberink JS. Mobile-bed effects in oscillatory sheet flow. J Geophys Res 2001;106(C11):27103-27115. 
\bibitem{Fredsoe} Fredsoe J. Modelling of non-cohesive sediment transport processes in the marine environment. Coast Eng 1993;21:71-103. 
\bibitem{Graf} Graf WH, Cellino M. Suspension flows in open channels: experimental study. J Hydraul Res 2002;40(4):435-447. 
\bibitem{Hsu} Hsu TW, Jan CD. Calibration of Businger–Arya type of eddy viscosity model's parameters. J Waterw Port Coastal Ocean Eng, ASCE 1998;124(5):281-284.
\bibitem{LeeHanes} Lee TH, Hanes DM. Comparison of field observations of the vertical distribution of suspended sand and its prediction by models, J Geophys Res 1996;101:3561–3572. 
\bibitem{McFetridge} McFetridge WF, Nielsen P. 1985. Sediment suspension by non-breaking waves over rippled beds. Technical Report No UFL/COEL-85/005, Coast Oc Eng Dept, University of Florida. 
\bibitem{Nakato} Nakato T, Locher FA, Glover JR, Kennedy JF. Wave entrainment of sediment from rippled beds. J Waterway Port Coastal Ocean Div 1977;103:83-99. 
\bibitem{Nielsen} Nielsen P. 1992. Coastal bottom boundary layers and sediment transport, World Sci., 324 p. 
\bibitem{NielsenPF} Nielsen P, Teakle IAL. Turbulent diffusion of momentum and suspended particles: A finite-mixing-length-theory. Phys Fluids 2004;16(7):2342-2348. 
\bibitem{Richardson} Richardson JF, Zaki WN. Sedimentation and fluidisation: part 1. Trans Inst Chem Eng 1954;32:35-53. 
\bibitem{Thorne} Thorne PD, Williams JJ, Davies AG. Suspended sediments under waves measured in a large-scale flume facility, J Geophys Res 2002;107(C8):3178. 
\bibitem{Thorne2} Thorne PD, Davies AG, Bell PS. Observations and analysis of sediment diffusivity profiles over sandy rippled beds under waves, J Geophys Res 2009;114:C02023. 
\bibitem{vanderWerf} van der Werf JJ, Ribberink JS, O'Donoghue T, Doucette JS. Modelling and measurement of sand transport processes over full-scale ripples in oscillatory flow, Coast Eng 2006;53(8):657-673. 

\end{thebibliography}
\end{document}